\documentclass[aps,pra,reprint,groupedaddress,showpacs]{revtex4-1}

\usepackage{graphicx}
\usepackage{amssymb,amsmath}
\usepackage{bm} 
\usepackage{bbold} 

\begin{document}

\title{Multichannel quantum-defect theory for magnetic Feshbach resonances 
	in heteronuclear group I systems}

\author{Constantinos Makrides}
\email[]{constantinos.makrides@rockets.utoledo.edu}
\author{Bo Gao}
\email[]{bo.gao@utoledo.edu}
\homepage[]{http://bgaowww.physics.utoledo.edu}
\affiliation{Department of Physics and Astronomy,
	University of Toledo, Mailstop 111,
	Toledo, Ohio 43606,
	USA}

\date{April 16, 2014}

\begin{abstract}

We present a multichannel quantum-defect theory for magnetic Feshbach resonances in the interaction of two heteronuclear group I atoms. The theory provides a unified and a uniform description of resonances in all partial waves, and enables the characterization of large number of resonances in terms of very few parameters. For the sample system of $^6$Li$^{40}$K, we present descriptions of all resonances in $aa$, $ab$, and $ba$ channels, in partial waves $s$ ($l=0$) through $h$ ($l=5$), and in a magnetic field of 0 through 1000 Gauss. All resonances, including those in nonzero partial waves, are fully characterized using the newly developed parametrization of Gao [Phy. Rev. A \textbf{84}, 022706 (2011)].

\end{abstract}


\maketitle

\section{Introduction}

Magnetic Feshbach resonance \cite{tie93} plays an important role in studies and applications of cold atoms (see \cite{chi10} and references therein). In the context of few-body and many-body physics, it is famously known for being the key mechanism for controlling and tuning of atomic interactions. In the context of more traditional atomic physics, it represents one of the most precise experimental measurements that can be carried out in the sub-thermal temperature regime and can be used to calibrate our understanding of atom-atom interactions (see, e.g., Refs.~\cite{Tiemann2009,kno11,Wang2013}).

Being intrinsically a multichannel phenomenon that often involves a large number of channels, a thorough understanding of magnetic Fesbach resonances, or more generally the understanding of atomic interaction in a magnetic field, remains a complex task with many open questions.  Around an $s$ wave resonance, the parametrization of the scattering length \cite{moe95,koh06,chi10}
\begin{equation}
a_{l=0}(B) = a_{\text{bg}l=0}\left(1-\frac{\Delta_{Bl=0}}{B-B_{0l=0}}\right) \;,
\label{eq:as1}
\end{equation}
in terms of three parameters: the resonance position  $B_{0l=0}$, the background scattering length $a_{\text{bg}l=0}$, and the width parameter $\Delta_{Bl=0}$, has greatly simplified the description of ultracold atomic interaction in a magnetic field and greatly facilitated its applications in cold-atom physics \cite{chi10}. For resonances in nonzero partial waves  (see, e.g., Refs.~\cite{chi00,mar02,reg03b,zha04,sch05,gae07,kno08,Wille2008,Wang2013}), however, similar parametrization has not existed until very recently \cite{gao11b}, due to the fact that they cannot be described using the standard effect range theory \cite{sch47,bla49,bet49}. Furthermore, even for an $s$ wave resonance, information contained in Eq.~(\ref{eq:as1}) is generally insufficient to fully characterize ultracold atomic interactions if the resonance is not of the ``broad'' type \cite{sto05,sim05,koh06,chi10,gao11b}.

In Ref.~\cite{gao11b}, we have derived a uniform parametrization of magnetic Feshbach resonances in arbitrary partial waves and analytic descriptions of ultracold atomic interaction around them. The theory is based on the multichannel quantum-defect theory (MQDT) of Ref.~\cite{gao05a} and on the QDT expansion of Ref.~\cite{gao09a}. We have shown that a magnetic Feshbach resonance in an arbitrary partial wave $l$, and the ultracold atomic interaction around it, can be fully characterized using five parameters. They can either be the set of $B_{0l}$, $K^{c0}_{\text{bg}l}$, $g_{\text{res}}$, $d_{Bl}$, and $s_E$ (or $C_6$), or the set of $B_{0l}$, $\widetilde{a}_{\text{bg}l}$, $\Delta_{Bl}$, $\delta\mu_l$ and $s_E$ (or $C_6$), with the latter set being the more direct generalization of the $s$ wave parametrization of Eq.~(\ref{eq:as1}) . Here $\widetilde{a}_{\text{bg}l}$ is a generalized background scattering length, well defined for all $l$, $\delta\mu_l$ is a differential magnetic moment, and $s_E$ is the characteristic energy scale associated with the $-C_6/R^6$ van der Waals interaction. These and other parameters are explained in detail in Ref.~\cite{gao11b} and in later sections. Used with the QDT expansion, or the generalized effective-range expansion derived from it \cite{gao09a,gao11b}, they give accurate analytic descriptions of atomic interactions around a magnetic Feshbach resonance, not only of the scattering properties, but also of the binding energies of a Feshbach molecule and of scattering at negative energies \cite{gao08a}. The parametrization works the same for both broad and narrow resonances, and resonances of intermediate characteristics.

This work gives a more detailed presentation of MQDT for heteronuclear atomic interaction in a magnetic field \cite{Burke1998,han09,gao11b,Ruzic2013}, with focus on its application in the description of Feshbach resonances, especially in the determination of resonance parameters mentioned above. Using the example of  $^6$Li$^{40}$K \cite{Wille2008,Tiemann2009,han09}, we show how MQDT provides an efficient and uniform description of large number of resonances, in different partial waves and over a wide range of fields, from the same few parameters that characterize atomic interaction in the absence of the field \cite{gao05a}. All resonances in $aa$, $ab$, and $ba$ channels, in partial waves $s$ ($l=0$) through $h$ ($l=5$), and in fields of 0 G through 1000 G, are fully characterized using the parameters introduced in Ref.~\cite{gao11b}.

\section{MQDT for atomic interaction in a magnetic field}

\subsection{Fragmentation channels for two heteronuclear group I atoms in a magnetic field}
\label{sec:fchannel}

\subsubsection{A single group I atom in a magnetic field}
\label{sec:alkaliB}

The Zeeman effect of a group I atom in a magnetic field $B$ is well known \cite{Breit1931}. We briefly summarize the result here with the intention of defining our notations.

A group I atom, with an electronic angular momentum of $J_A=1/2$ and a nuclear spin of $I_A$ in a magnetic field has $2(2I_A+1)$ states that are fully split. They are well labeled by $|F_A,m_f\rangle$ in a weak field, where $\bm{F}_A=\bm{J}_A+\bm{I}_A$, and by $|J_A,m_{JA};I_A,m_{IA}\rangle$ in a strong field. For intermediate fields, both labeling can in principle still be used to uniquely identify a state, but are no longer ideal since quantum numbers other than $m_f=m_{JA}+m_{IA}$ are generally no longer good quantum numbers.

\begin{figure}
\includegraphics[width=\columnwidth]{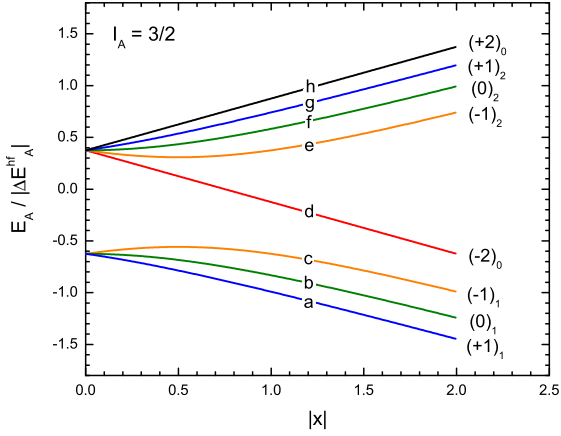}
\caption{(Color online) Illustration of the energy diagram for a group I atom with $I_A=3/2$ in a magnetic field. It applies to $^7$Li, $^{23}$Na, $^{39}$K, and $^{87}$Rb, upon ignoring the small correction due to $g_I$. Here $x = (g_J-g_I)\mu_BB/\Delta E^{\text{hf}}_A$ is a scaled magnetic field. Both the concise alphabetic labeling and the more detailed labeling of states are illustrated.
\label{fig:Rb87hfsB}}
\end{figure}
\begin{figure}
\includegraphics[width=\columnwidth]{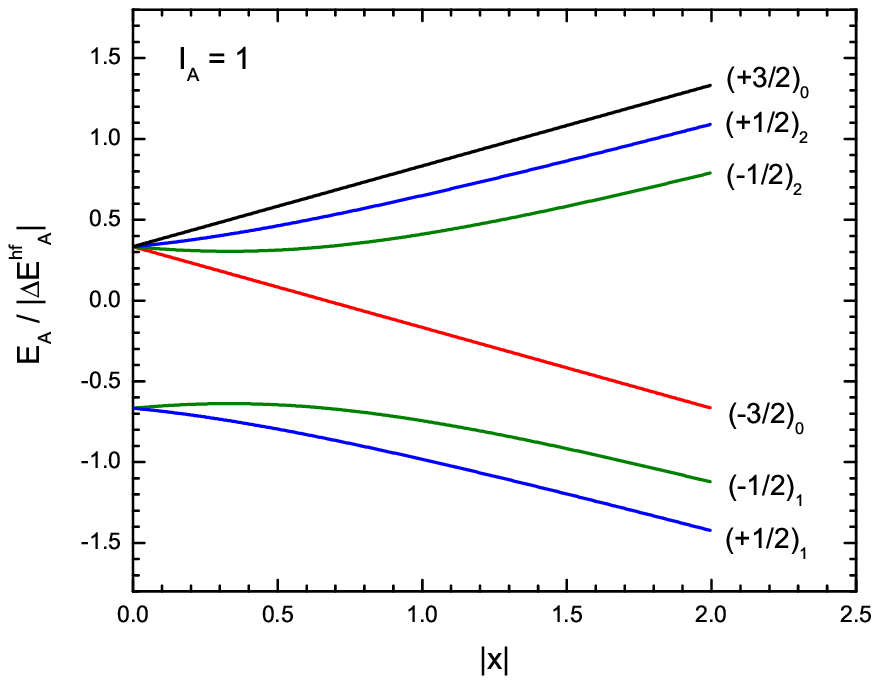}
\caption{(Color online) Illustration of the energy diagram for a group I atom with $I_A=1$ in a magnetic field, applicable to $^{6}$Li.
\label{fig:Li6hfsB}}
\end{figure}
\begin{figure}
\includegraphics[width=\columnwidth]{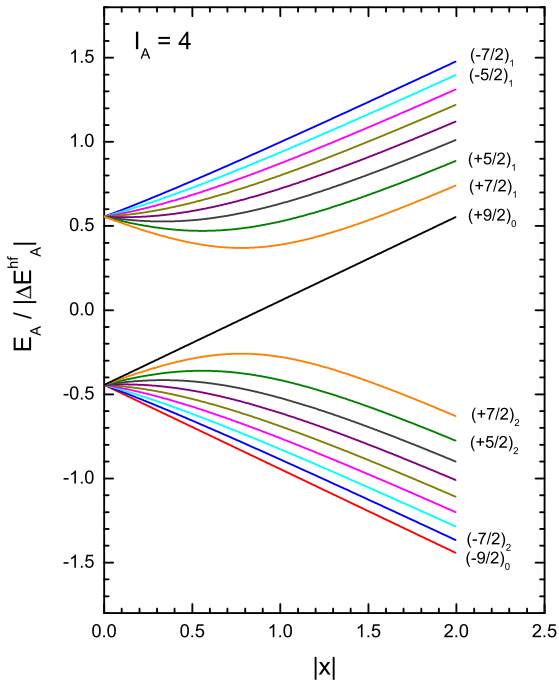}
\caption{(Color online) Illustration of the energy diagram for a group I atom with $I_A=4$ in a magnetic field, applicable to  $^{40}$K.
\label{fig:K40hfsB}}
\end{figure}

We will identify the states using the notation of $|(m_f)_\alpha\rangle$, to emphasize that if one is interested in a wide range of magnetic fields, only $m_f$ is a good quantum number, whereas different $F_A$ states are generally coupled. Specifically, in the single-atom $J_AI_A$ basis, with basis states $|J_A,m_{JA};I_A,m_{IA}\rangle$ to be denoted by a simpler notation of $|m_{JA};m_{IA}\rangle$, the states of a group I atom in a magnetic field $B$ are given, for $|m_f|=I_A+1/2$, by
\begin{equation}
|(I_A+1/2)_0\rangle = \left|1/2;I_A\right\rangle \;,
\label{eq:Bstate1}
\end{equation}
\begin{equation}
|(-I_A-1/2)_0\rangle = \left|-1/2;-I_A\right\rangle \;.
\end{equation}
For $m_f=-I_A+1/2,\cdots,I_A-1/2$, there are two states corresponding to each $m_f$. They are distinguished by $\alpha=1,2$, and are given by
\begin{eqnarray}
|(m_f)_1\rangle &=& \sin(\theta^B/2)\left|1/2;m_f-1/2\right\rangle \nonumber\\
	& &-\cos(\theta^B/2)\left|-1/2;m_f+1/2\right\rangle\;, \\
|(m_f)_2\rangle &=& \cos(\theta^B/2)\left|1/2;m_f-1/2\right\rangle \nonumber\\
	& &+\sin(\theta^B/2)\left|-1/2;m_f+1/2\right\rangle\;.
\label{eq:Bstate4}
\end{eqnarray}
Here $0\le\theta^B(m_f)<\pi$ and
\begin{equation}
\tan\theta^B(m_f) = \frac{\sqrt{1-\left(\frac{2m_f}{2I_A+1}\right)^2}}
	{\frac{2m_f}{2I_A+1}+x} \;,
\end{equation}
and we have defined
$x = (g_J-g_I)\mu_BB/\Delta E^{\text{hf}}_A$, in which $\mu_B$ is the Bohr magneton and $g_J$ and $g_I$ are the gyromagnetic ratios for the electron and the nucleus, respectively \cite{Arimondo1977}. The hyperfine splitting $\Delta E^{\text{hf}}_A$ is defined as $\Delta E^{\text{hf}}_A\equiv E_A(F_A=I_A+1/2)-E(F_A=I_A-1/2)$ in the absence of magnetic field. It is positive for most alkali-metal atoms, but negative for $^{40}K$. In defining states,   we have chosen a phase convention such that
\begin{align*}
|(m_f)_1\rangle \stackrel{B\rightarrow 0}{\sim} & |F_A=I_A-1/2,m_f\rangle \;,\\
|(m_f)_2\rangle \stackrel{B\rightarrow 0}{\sim} & |F_A=I_A+1/2,m_f\rangle \;.
\end{align*}
The energies of the states $|(m_f)_\alpha\rangle$ are given, for $|m_f|=I_A+1/2$, by
\begin{equation}
E_A((I_A+1/2)_0) = \Delta E^{\text{hf}}_A\left[\frac{I_A}{2I_A+1}
	+\frac{1}{2}\frac{g_J+2I_Ag_I}{g_J-g_I}x\right]\;,
\end{equation}
\begin{equation}
E_A((-I_A-1/2)_0) = \Delta E^{\text{hf}}_A\left[\frac{I_A}{2I_A+1}
	-\frac{1}{2}\frac{g_J+2I_Ag_I}{g_J-g_I}x\right] \;,
\end{equation}
and for other states, $m_f=-I_A+1/2,\cdots,I_A-1/2$, by \cite{Breit1931}
\begin{eqnarray}
E_A((m_f)_1) &=& \frac{1}{2}\Delta E^{\text{hf}}_A\left[-\frac{1}{2I_A+1}
	+\frac{2m_fg_I}{g_J-g_I}x \right.\nonumber\\
	& &-\left.\left(1+\frac{4m_f}{2I_A+1}x+x^2\right)^{1/2}\right] \;,
\end{eqnarray}
\begin{eqnarray}
E_A((m_f)_2) &=& \frac{1}{2}\Delta E^{\text{hf}}_A\left[-\frac{1}{2I_A+1}
	+\frac{2m_fg_I}{g_J-g_I}x \right.\nonumber\\
	& &+\left.\left(1+\frac{4m_f}{2I_A+1}x+x^2\right)^{1/2}\right] \;.
\end{eqnarray}
For $\Delta E^{\text{hf}}_A>0$, the state $(m_f)_2$ has higher enery
than $(m_f)_1$. The reverse is true for $\Delta E^{\text{hf}}_A<0$
	
The other common labeling of states of a group I atom in a magnetic field is alphabetic ordering, $a$, $b$, $c$, \dots, in the order of increasing energy \cite{chi10}. This labeling is very concise and can be convenient, but is, at times, not sufficiently informative. We will utilize both set of notations, alphabetically for convenience and for comparison with earlier works, and the $|(m_f)_\alpha\rangle$ notation when necessary.

Figure~\ref{fig:Rb87hfsB} illustrates the energy diagram for alkali-metal atoms $^7$Li, $^{23}$Na, $^{39}$K, and $^{87}$Rb, all with $I_A=3/2$. It shows that for $\Delta E^{\text{hf}}_A>0$, the lowest atomic state, the $a$ state, is always $|(I_A-1/2)_1\rangle$, and the $b$ state is always $|(I_A-3/2)_1\rangle$. Figure~\ref{fig:Li6hfsB} depicts the energy diagram for a $^6$Li atom, which has an $I_A=1$. Figure~\ref{fig:K40hfsB} depicts the energy diagram for a $^{40}$K atom, which has an $I_A=4$. It illustrates that for $\Delta E^{\text{hf}}_A<0$, the lowest atomic state, the $a$ state, is always $|(-I_A-1/2)_0\rangle$, and the $b$ state is always $|(-I_A+1/2)_2\rangle$.

Equations~(\ref{eq:Bstate1})-(\ref{eq:Bstate4}) define a unitary transformation $U^B_A$, with elements $\langle m_{JA};m_{IA}|(m_f)_\alpha\rangle$, that relates the single atom $J_AI_A$ basis to the single-atom states in a magnetic field, which diagonalize both the hyperfine and the magnetic interactions. It is a one-dimensional unit matrix for $|m_f|=I_A+1/2$, and is given by
\begin{equation}
U^B_A =
\left(
\begin{array}{lr}
\sin(\theta^B/2) & -\cos(\theta^B/2) \\
\cos(\theta^B/2) & \sin(\theta^B/2) 
\end{array}
\right) \;,
\end{equation}
for $m_f=-I_A+1/2,\cdots,I_A-1/2$.

\subsubsection{Two heteronuclear group I atoms in a magnetic field}

The fragmentation channels for atomic interaction in a magnetic field are determined by two-atom states in a magnetic field. For two heteronuclear group I atoms, labeled by 1 and 2, in a magnetic field, there are $2(2I_1+1)\times 2(2I_2+1)$ number of channels for each partial wave $l$. They can each be identified by $|(m_{f1})_{\alpha_1},(m_{f2})_{\alpha_2},l\rangle$, with $|(m_{fx})_{\alpha_x}\rangle$ characterizing the state of atom $x$ in a magnetic field in a way as specified in Sec.~\ref{sec:alkaliB}. The channel threshold energies are given by $E_1((m_{f1})_{\alpha_1})+E_2((m_{f2})_{\alpha_2})$.

Ignoring the weak magnetic dipole-dipole  \cite{sto88,moe95} and second-order spin-orbit interactions \cite{mie96,kot00,leo00}, $M_F = m_{f1}+m_{f2}$ is conserved. The interaction and the scattering matrices are block-diagonal with each block labeled by $M_F$. The number of (coupled) channels in each block is independent of $l$, and is determined by $|M_F|$. For instance, $M_F = \pm(I_1+I_2+1)$ blocks both have only a single channel, $M_F = \pm(I_1+I_1)$ blocks both have 4  channels, and blocks with $|I_1-I_2|<|M_F|<I_1+I_2$ have $4(I_1+I_2+1-|M_F|)$ channels, \textit{etc}. The total number of channels for each $l$ adds up to $2(2I_1+1)\times 2(2I_2+1)$.

\subsection{MQDT}
\label{sec:MQDT}

MQDT for atomic interaction in a magnetic field \cite{Burke1998,han09,gao11b,Ruzic2013}, to be deployed here, is formally the same as MQDT for atomic interaction in the absence of external fields \cite{gao05a}. The theory takes full advantage of the physics that both the energy dependence \cite{gao98b} and the partial wave dependence \cite{gao01} of the atomic interaction around a threshold are dominated by effects of the long-range potential, which are described by in a set of universal QDT functions \cite{gao98a,gao08a}. The short-range contribution is isolated to a short-range $K^c$ matrix that is insensitive to both the energy and the partial wave.

For an $N$-channel problem and at energies where all channels are open, MQDT gives the physical $K$ matrix, in our case the $K^{M_Fl}$, as \cite{gao05a}
\begin{equation}
{ K^{M_Fl}} = -( { Z}^c_{fc}-{ Z}^c_{gc}{ K}^{c} )({ Z}^c_{fs} - { Z}^c_{gs}{ K}^{c})^{-1} \;,
\label{eq:Kphyo}
\end{equation}
where $Z^c_{xy}$s are $N\times N$ diagonal matrices with elements $Z^c_{xy}(\epsilon_{si},l)$ being the $Z^c_{xy}$ functions \cite{gao08a} evaluated at scaled energies $\epsilon_{si}=(E-E_i)/s_E$ relative to the respective channel threshold $E_i$. Here $s_E = (\hbar^2/2\mu)(1/\beta_6)^2$ and $\beta_6 = (2\mu C_6/\hbar^2)^{1/4}$ are the characteristic energy and length scales, respectively, associated with the $-C_6/R^6$ van der Waals potential, with $\mu$ being the reduced mass. At energies where $N_o$ channels are open, and $N_c=N-N_o$ channels are closed, MQDT gives \cite{gao05a}
\begin{equation} 
K^{M_Fl} = -( Z^c_{fc}-Z^c_{gc}K^c_{\mathrm{eff}} )(Z^c_{fs} - Z^c_{gs}K^c_{\mathrm{eff}})^{-1} \;,
\label{eq:Kphy}
\end{equation}
where
\begin{equation}
K^c_{\mathrm{eff}} = K^{c}_{oo}+K^{c}_{oc}(\chi^c -K^{c}_{cc})^{-1}K^{c}_{co} \;,
\label{eq:Kceff}
\end{equation}
in which $\chi^c$ is a $N_c\times N_c$ diagonal matrix with elements $\chi^c_l(\epsilon_{si},l)$ \cite{gao08a}, and $K^{c}_{oo}$, $K^{c}_{oc}$, $K^{c}_{co}$, and $K^{c}_{cc}$, are submatrices of $K^{c}$ corresponding to open-open, open-closed, closed-open, and closed-closed channels, respectively. Equation~(\ref{eq:Kphy}) is formally the same as Eq.~(\ref{eq:Kphyo}), except that the $K^c$ matrix is replaced by $K^c_{\mathrm{eff}}$ that accounts for the effects of closed channels. From the physical $K$ matrix, the physical $S$ matrix is obtained from \cite{gao96}
\begin{equation}
{S^{M_Fl}} = (\mathbb{1}+i{K^{M_Fl}})(\mathbb{1}-i{K^{M_Fl}})^{-1} \;, 
\end{equation}
from which all the scattering properties can be determined.
At energies where all channels are closed, the bound spectrum can be determined either from
\begin{equation}
\det[\chi^c(E) - K^{c}]=0 \;,
\label{eq:bound}
\end{equation}
where $K^c$ is the full $N\times N$ $K^c$ matrix, or from equivalent effective single-channel or effective multichannel problems, as discussed in Ref.~\cite{gao11b}.
 
The presence of a magnetic field changes the channel structure for atomic interaction, including both the channel wave functions and the channel threshold energies. It is though these changes that atomic interaction depends on the magnetic field. The change of channel threshold energies is as discussed in Sec.~\ref{sec:fchannel}. The change of  channel wave functions leads to a field-dependent $K^c$ matrix. This aspect is discussed in the following subsection.

\subsection{The $K^c$ matrix for heteronuclear group I systems}
\label{sec:Kc}

The short-range $K^c$ matrix for two heteronuclear group I atoms in a magnetic field can be obtained through a frame transformation from two single-channel $K^c$s: $K^c_S(\epsilon,l)$ for spin singlet $^1\Sigma^+_g$ state, and $K^c_T(\epsilon,l)$ for spin triplet $^3\Sigma^+_u$ state, which are the same parameters describing atomic interaction in the absence of the magnetic field \cite{gao05a}.
Specifically,  the $K^c$ matrix, being a short-range $K$ matrix, has much simpler representation in condensation channels that diagonalize the short-range interaction \cite{gao96,gao05a}. For interaction of group I atoms, a convenient choice of condensation channels is the $JI$ coupled basis, in which a state is labeled by $|(JM_J,IM_I)l\rangle$ where $\bm{J}=\bm{J}_1+\bm{J}_2$ and $\bm{I}=\bm{I}_1+\bm{I}_2$.
The short-range $K^c$ matrix is to an excellent approximation diagonal in this basis with elements given by
\begin{multline}
\langle(J'M'_J,I'M'_I)l| K^{c(JI)}|(JM_J,IM_I)l\rangle \\
	=K^c_J(\epsilon,l)\delta_{J'J}\delta_{M'_JM_J}\delta_{I'I}\delta_{M'_IM_I} \;,
\end{multline}
where $K^c_{J=0}$ corresponds to $K^c_S$, and $K^c_{J=1}$ corresponds to $K^c_T$ \cite{gao05a}. 

From $K^{c(JI)}$, the $K^c$ matrix in fragmentation channels in a magnetic field, as defined in Sec.~\ref{sec:fchannel},
is given by
\begin{equation}
K^c(B) = U^{B\dagger}K^{c(JI)}U^{B} \;,
\label{eq:Kctrans}
\end{equation}
where
\begin{equation}
U^B = U^{JI}(U^B_1\otimes U^B_2) \;.
\end{equation}
Here $U^B_1$ and $U^B_2$ are unitary transformations, discussed earlier, describing the re-coupling of atom 1 states and atom 2 states by the magnetic field, respectively. $U^{JI}$ is a unitary transformation with elements, $\langle JM_J|J_1m_{J1},J_2m_{J2}\rangle\langle IM_I|I_1m_{I1},I_2m_{I2}\rangle$,  given by the Clebsch-Gordon coefficients describing the coupling of $\bm{J}=\bm{J}_1+\bm{J}_2$ and  $\bm{I}=\bm{I}_1+\bm{I}_2$. 

The applicability of frame transformation depends critically on $K^c$ being a short-range matrix \cite{gao05a}. It is not applicable to the physical $K$ matrix except at energies much greater than hyperfine and magnetic splittings.  The $K^c$ matrix, when regarded as an operator associated with short-range atomic interaction, is dominated by electrostatic and exchange interactions, which, to an excellent approximation, are not affected by external fields and other weak interactions such as hyperfine interaction. External fields and hyperfine interaction only change the basis in which $K^c$ needs to be represented, not the operator itself.

From the unitarity of $U^B$, one can show rigorously that all off-diagonal elements of $K^c(B)$ are proportional to $\Delta K^c\equiv K^c_S-K^c_T$. It is a general feature applicable both in and in the absence of the magnetic field \cite{gao05a}. It implies that the coupling of different fragmentation channels, which is responsible both for the existence of Feshbach resonances and for inelastic scattering,  is due primarily to the difference between the singlet and triplet scattering, as reflected, e.g., in the difference in $s$ wave scattering lengths $\Delta a_{l=0}\equiv a^S_{l=0}-a^T_{l=0}$, where $a^S_{l=0}$ and $a^T_{l=0}$ are the $s$ wave scattering lengths for the singlet and the triplet state, respectively. The coupling due to magnetic dipole-dipole  \cite{sto88,moe95} and second-order spin-orbit interactions \cite{mie96,kot00,leo00} are much weaker effects \cite{kno11,Wang2013,Ruzic2013}, ignored in the present study.

\section{MQDT for magnetic Feshbach resonances}

The MQDT formulation of the previous section provides a systematic understanding  of atomic interaction in a magnetic field, including both elastic and inelastic processes, and over a wide range of energies and fields. It works the same for all heteronuclear group I atoms, and uses the same parameters, $K^c_S(\epsilon,l)$ and $K^c_T(\epsilon,l)$, that one would use for interaction in the absence of the field \cite{gao05a}. We focus here on the application of the theory to (zero-energy) magnetic Feshbach resonances. Other applications will be addressed elsewhere. 

We illustrate the theory with sample results for $^6$Li$^{40}$K. To set the benchmark for future investigations and to keep the focus on concepts, we further limit ourselves to  baseline MQDT results that ignores the energy and the partial-wave dependences of the short-range parameters, namely in the approximation of $K^c_S(\epsilon,l)\approx K^c_S(\epsilon=0,l=0)$ and $K^c_T(\epsilon,l)\approx K^c_T(\epsilon=0,l=0)$. In this baseline description, all aspects of cold atomic interaction, including parameters for all magnetic Feshbach resonances in all partial waves, are determined from three parameters \cite{gao05a,han09}: the $C_6$ coefficient, the singlet $s$ wave scattering length $a^S_{l=0}$, and the triplet $s$ wave scattering length $a^T_{l=0}$, in addition to well known atomic parameters such as the atomic mass and hyperfine splitting.

$^6$Li$^{40}$K is an interesting quantum system of two fermionic atoms. The magnetic Feshbach resonances have been investigated experimentally by Wille \textit{et al.}  \cite{Wille2008}, and theoretically by Hanna \textit{et al.}  \cite{han09} and by Tiemann  \textit{et al.}  \cite{Tiemann2009}. $^6$Li has $I_1=1$ with a hyperfine splitting of $\Delta E^{\mathrm{hf}}_1/h = 228.205$ MHz and $g_{I1} = -4.477 \times 10^{-4}$ \cite{Arimondo1977}. $^{40}$K has $I_2=4$ with a hyperfine splitting of $\Delta E^{\mathrm{hf}}_2/h = -1285.79$ MHz and $g_{I2} = 1.765 \times 10^{-4}$ \cite{Arimondo1977}.
The three parameters used in our baseline MQDT description are $C_6 = 2322$ a.u. \cite{Derevianko2001,Tiemann2009}, $a^S_{l=0}= 52.50$ a.u., and $a^T_{l=0}= 63.70$ a.u. The singlet and the triplet $s$ wave scattering lengths are slightly adjusted from the values of Hanna \textit{et al.}  \cite{han09} for a better agreement with experimental $s$ wave Feshbach resonance positions at low fields.  In terms of quantum defects \cite{gao08a}, the adopted scattering lengths correspond to  $\mu^c_S(\epsilon=0,l=0) = 0.3938$ for the singlet state $\mu^c_T(\epsilon=0,l=0) = 0.3202$ for the triplet state.

The $s$ wave scattering lengths give $K^c_S(\epsilon,l)\approx K^c_S(\epsilon=0,l=0) = -16.91$ and $K^c_T(\epsilon,l)\approx K^c_T(\epsilon=0,l=0) = 5.748$ \cite{gao05a,gao08a}, from which the $K^c$ matrix in a magnetic field is constructed as discussed in Sec.~\ref{sec:Kc}. From the $K^c$ matrix, the MQDT of Sec.~\ref{sec:MQDT} provides a complete description of cold atomic interactions in a $B$ field. The $C_6$ coefficient gives a length scale of $\beta_6 = 81.56$ a.u. and a corresponding energy scale of $s_E/k_B = 2.490$ mK for $^6$Li-$^{40}$K van der Waals interaction.
 

\begin{figure}[!]
\includegraphics[width=\columnwidth]{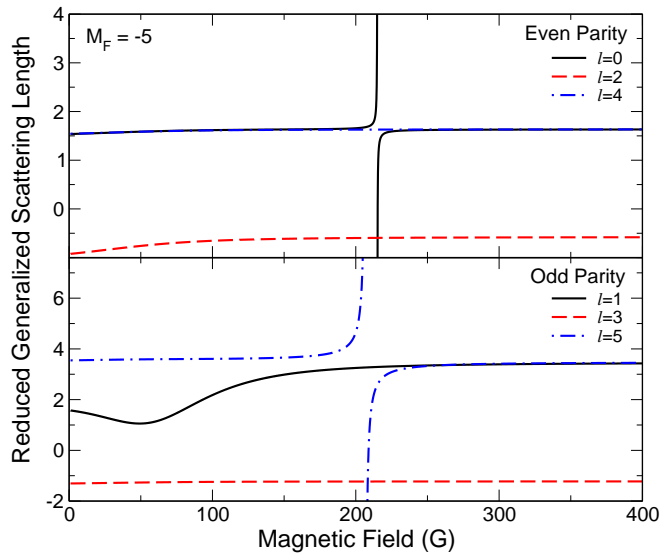}
\caption{(Color online) Reduced generalized scattering lengths,
$\widetilde{a}_{l}(B)/\bar{a}_{l}$,
versus magnetic field for $^6$Li$^{40}$K in $ba$
channel (corresponding to the manifold of $M_F=-5$),
for partial waves $s$ through $h$.
\label{fig:Li6K40mf-5}}
\end{figure}

\begin{figure}[!]
\includegraphics[width=\columnwidth]{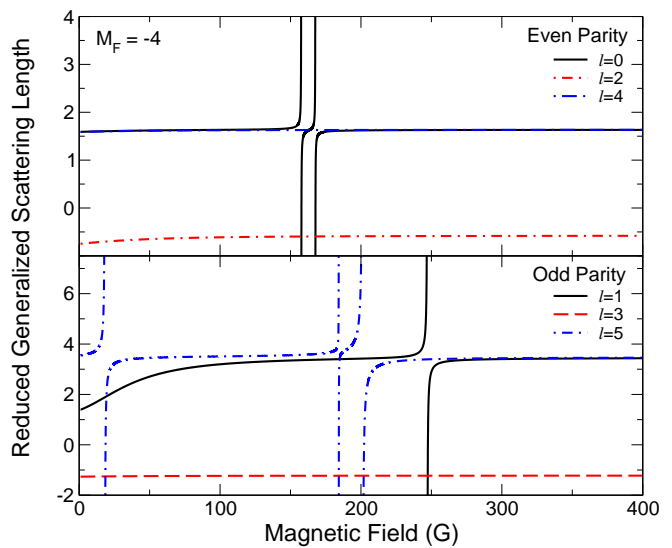}
\caption{(Color online) Reduced generalized scattering lengths,
$\widetilde{a}_{l}(B)/\bar{a}_{l}$,
versus magnetic field for $^6$Li$^{40}$K in $aa$
channel (corresponding to the manifold of $M_F=-4$),
for partial waves $s$ through $h$.
\label{fig:Li6K40mf-4}}
\end{figure}

\begin{figure}[!]
\includegraphics[width=\columnwidth]{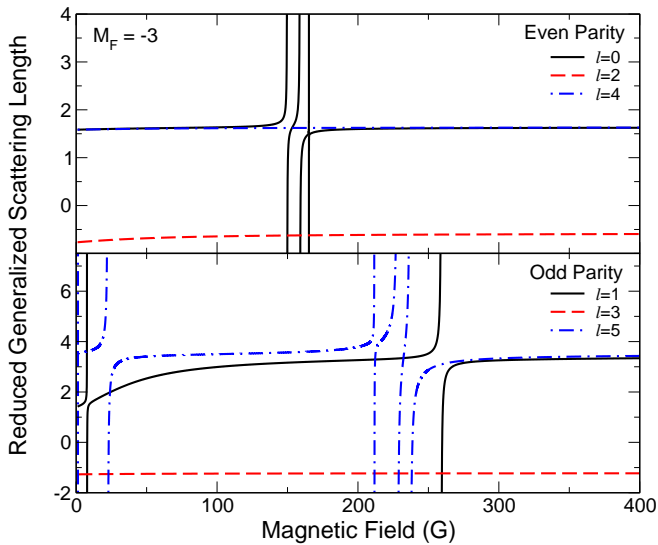}
\caption{(Color online) Reduced generalized scattering lengths,
$\widetilde{a}_{l}(B)/\bar{a}_{l}$,
versus magnetic field for $^6$Li$^{40}$K in $ab$
channel (corresponding to the manifold of $M_F=-3$),
for partial waves $s$ through $h$.
\label{fig:Li6K40mf-3}}
\end{figure}

In understanding magnetic Feshbach resonances, the focus is on atomic interaction near the lowest threshold for each $M_F$. For all energies below the second lowest threshold, both scattering and bound states can be described by an effective single-channel problem with  \cite{gao11b}
\begin{equation}
K^c_{\text{eff}} = K^{c}_{11}+K^{c}_{1c}(\chi^c - K^{c}_{cc})^{-1}K^{c}_{c1} \;,
\label{eq:Kceff}
\end{equation}
or with its corresponding $K^{c0}_l$ matrix  \cite{gao11b}
\begin{equation}
K^{c0}_l(\epsilon,B) = \frac{K^c_{\text{eff}}(\epsilon, l)-\tan(\pi\nu_0/2)}
	{1+\tan(\pi\nu_0/2)K^c_{\text{eff}}(\epsilon, l)} \;.
\label{eq:Kc0l}
\end{equation}
In Eq.~(\ref{eq:Kceff}), ``1'' refers to the lowest-energy channel for the $M_F$ manifold, and ``c'' refers to all other (closed) channels. In Eq.~(\ref{eq:Kc0l}), $\nu_0 = (2l+1)/4$, and $\epsilon=E-E_1$ is the energy relative to the lowest threshold.

In terms of $K^c_{\text{eff}}$, scattering below the second threshold, where only elastic scattering is possible, is described by the physical $K$ matrix
\begin{equation}
\tan\delta_l = ( Z^{c}_{gc}K^{c}_{\text{eff}}-Z^{c}_{fc} )
	(Z^{c}_{fs} - Z^{c}_{gs}K^{c}_{\text{eff}})^{-1} \;,
\label{eq:Kphy}	
\end{equation}
from which the elastic scattering cross section can be calculated in the standard manner.

In terms of the value of $K^{c0}_l(\epsilon,B)$ at zero energy (namely at the lowest threshold), defined by
\begin{equation}
K^{c0}_{l}(B)\equiv K^{c0}_{l}(\epsilon=0,B) \;,
\end{equation}
the resonance locations, $B_{0l}$, of zero-energy magnetic Feshbach resonances in any partial wave $l$, can be determined by the roots of $K^{c0}_{l}(B)$, namely as the solutions of  \cite{gao11b}
\begin{equation}
K^{c0}_{l}(B) = 0 \;.
\label{eq:B0l}
\end{equation}
The generalized scattering length as a function of magnetic field, for an arbitrary partial wave $l$, is given by \cite{gao11b}
\begin{equation}
\widetilde{a}_{l}(B) = \bar{a}_{l}\left((-1)^l+\frac{1}{K^{c0}_{l}(B)}\right) \;,
\label{eq:ga} 
\end{equation}
where $\bar{a}_{l}=\bar{a}_{sl}\beta_6^{2l+1}$ is the mean scattering length (with scale included) with
\begin{equation}
\bar{a}_{sl} = \frac{\pi^2}{2^{4l+1}[\Gamma(l/2+1/4)\Gamma(l+3/2)]^2} \;,
\end{equation}
being the scaled mean scattering length for partial wave $l$ \cite{gao09a}. 

\begin{table*}[!]
\caption{Selective magnetic Feshbach resonances for the $^6$Li-$^{40}$K system that have been found experimentally \cite{Wille2008}, and investigated in an earlier MQDT theory \cite{han09}. Before the work of Ref.~\cite{gao11b}, the width parameter $\Delta_{Bl}$ was not rigorously defined for nonzero partial waves, and was therefore not provided in the previous theory \cite{han09} for $p$ wave resonances.}
\begin{ruledtabular}
\begin{tabular}{c c c||c| c|| c | c|| c}
\multicolumn{3}{c||}{Identification}& \multicolumn{2}{c||}{Present MQDT} & \multicolumn{2}{c||}{Previous MQDT \cite{han09}} & Experiment   \cite{Wille2008}\\ \hline \hline
Channel & $M_f$ & $l$ & $B_{0l}$ (G) & $\Delta_{Bl}$ (G) & $B_{0l}$ (G) & $\Delta_{Bl}$ (G) & $B_{0l}$ (G)\\ \hline
$ba$ & -5 & 0 & 215.0 & 0.245 & 213.6 & 0.28 & 215.6\\ \hline
$aa$ & -4 & 0 & 157.6 & 0.141 & 159.3 & 0.22 & 157.6\\ \hline
$aa$ & -4 & 0 & 167.6 & 0.107 & 170.1 & 0.07 & 168.2\\ \hline
$aa$ & -4 & 1 & 247.1 & 0.447 & 258.0 & - & 249.0\\ \hline
$ab$ & -3 & 0 & 149.7 & 0.242 & 153.1 & 0.42 & 149.2\\ \hline
$ab$ & -3 & 0 & 158.8 & 0.425 & 159.6 & 0.16 & 159.5\\ \hline
$ab$ & -3 & 1 & 259.1 & 0.692 & 260.8 & - & 263.0\\
\end{tabular}
\end{ruledtabular}
\label{tab:LiKRes}
\end{table*}

\begin{table*}
\caption{A complete list of  magnetic Feshbach resonances and their parameters for $^6$Li-$^{40}$K system in $ba$, $aa$, and $ab$ channels, in partial wave $s$ ($l=0$) through $h$ ($l=5$), and in magnetic fields from 0 through 1009 G. A large number of resonances are predicted in the $h$ wave ($l=5$). } 
\begin{ruledtabular}
\begin{tabular}{c c c c c c c c c c c}
Channel& $M_f$ & $l$ & $B_{0l}$ (G) & $\Delta_{B}$ (G) & $\widetilde{a}_{\text{bg}l}/\bar{a}_{sl}$ & $\delta \mu_l / \mu_B$ & $K^{c0}_{\text{bg}l}$ & $g_{\text{res}}$ & $\zeta_{\text{res}}$  & $d_{Bl}$ (G) \\ \hline 
$ba$ & -5 & 0 & 215.0 & 0.245  & 1.631 & 1.776 & 1.586 & -0.03036 &  0.006383 & -0.6336 \\ \hline 
$ba$ & -5 & 5 & 206.7 & 2.189  & 3.509 & 1.769 & 0.222 & -0.08129 & -0.003133 & -1.703  \\ \hline 
$aa$ & -4 & 0 & 157.6 & 0.141  & 1.648 & 1.660 & 1.543 & -0.01605 &  0.003467 & -0.3583 \\ \hline 
$aa$ & -4 & 0 & 167.6 & 0.107  & 1.608 & 1.800 & 1.644 & -0.01379 &  0.002795 & -0.2839 \\ \hline 
$aa$ & -4 & 1 & 247.1 & 0.447  & 3.415 & 0.152 & 0.227 & -0.00142 & -0.001255 & -0.3460 \\ \hline 
$aa$ & -4 & 5 & 18.22 & 0.506  & 3.477 & 2.854 & 0.223 & -0.03027 & -0.001158 & -0.3932 \\ \hline 
$aa$ & -4 & 5 & 184.3 & 0.053  & 3.705 & 1.678 & 0.213 & -0.00190 & -0.000077 & -0.0420 \\ \hline 
$aa$ & -4 & 5 & 201.1 & 1.100  & 3.468 & 1.830 & 0.224 & -0.04213 & -0.001609 & -0.8535 \\ \hline 
$aa$ & -4 & 5 & 981.7 & 0.028  & 3.578 & 1.890 & 0.218 & -0.00111 & -0.000043 & -0.0217 \\ \hline 
$aa$ & -4 & 5 & 999.8 & 0.596  & 3.458 & 1.956 & 0.224 & -0.02440 & -0.000930 & -0.4624 \\ \hline 
$ab$ & -3 & 0 & 149.7 & 0.242  & 1.697 & 1.588 & 1.435 & -0.02528 &  0.005874 & -0.5901 \\ \hline 
$ab$ & -3 & 0 & 158.8 & 0.425  & 1.579 & 1.758 & 1.726 & -0.05489 &  0.01060  & -1.157  \\ \hline 
$ab$ & -3 & 0 & 165.2 & 0.004  & 1.490 & 1.815 & 2.043 & -0.00059 &  0.000096 & -0.0121 \\ \hline 
$ab$ & -3 & 1 & 7.721 & 0.092  & 1.557 & 1.005 & 0.391 & -0.00152 & -0.000776 & -0.05597\\ \hline 
$ab$ & -3 & 1 & 259.1 & 0.692  & 3.284 & 0.163 & 0.233 & -0.00234 & -0.002008 & -0.5308 \\ \hline 
$ab$ & -3 & 5 & 1.265 & 0.0004 & 3.551 & 2.308 & 0.220 & -0.00002 & -0.000001 & -0.00034\\ \hline 
$ab$ & -3 & 5 & 22.43 & 0.670  & 3.466 & 2.372 & 0.224 & -0.03329 & -0.001271 & -0.5203 \\ \hline 
$ab$ & -3 & 5 & 211.7 & 0.088  & 3.942 & 1.510 & 0.202 & -0.00285 & -0.000121 & -0.07011\\ \hline 
$ab$ & -3 & 5 & 228.1 & 1.155  & 4.100 & 1.672 & 0.196 & -0.04189 & -0.001826 & -0.9286 \\ \hline 
$ab$ & -3 & 5 & 237.1 & 1.593  & 3.051 & 1.760 & 0.247 & -0.05696 & -0.001972 & -1.199  \\ \hline 
$ab$ & -3 & 5 & 1009  & 1.103  & 4.384 & 1.901 & 0.186 & -0.04604 & -0.002118 & -0.8978 \\ \hline 
\end{tabular}
\end{ruledtabular}
\label{tab:LiKAllParam}
\end{table*}

This MQDT-based formalism provides a unified and a uniform understanding of magnetic Feshbach resonances in all partial waves, as illustrated in Figs.~\ref{fig:Li6K40mf-5}-\ref{fig:Li6K40mf-3} for $^6$Li-$^{40}$K. Figure~\ref{fig:Li6K40mf-5} depicts the reduced generalized scattering lengths, $\widetilde{a}_{l}(B)/\bar{a}_{l}$, versus magnetic field for $^6$Li$^{40}$K in the $ba$ channel, for all partial waves $s$ ($l=0$) through $h$ ($l=5$). Here $ba$ refers to $^6$Li in $b$ state, namely the $|(-1/2)_1\rangle$ state (c.f.~Fig.~\ref{fig:Li6hfsB}), and $^{40}$K in $a$ state, namely the $|(-9/2)_0\rangle$ state (c.f.~Fig.~\ref{fig:K40hfsB}). It is the lowest energy channel of the $M_F=-5$ manifold with 4 coupled channels. We note that by plotting the dimensionless reduced generalized scattering lengths, the understanding of different partial waves are put on the same footing and can be depicted in the same figure. The plots are grouped into an even parity group and an odd parity group, as they are not coupled and can in principle be probed independently. Figures~\ref{fig:Li6K40mf-4} and \ref{fig:Li6K40mf-3} show similar results for the $aa$  channel ($M_F=-4$ manifold with 8 coupled channels) and the $ab$ channel ($M_F=-3$ manifold with 11 coupled channels), respectively. 

The resonances positions, determined from Eq.~(\ref{eq:B0l}), are part of the parameters tabulated in Tables~\ref{tab:LiKRes} and \ref{tab:LiKAllParam}.
Table~\ref{tab:LiKRes} provides a subset of our results for magnetic Feshbach resonances that have been measured experimentally \cite{Wille2008} and investigated in an earlier MQDT study \cite{han09}. The resonance positions, which are the only parameters measured, are found to be consistent with previous theory \cite{han09} and in good agreement with experiment \cite{Wille2008}.

Around each resonance, the dependence of the generalized scattering length on the magnetic field can be characterized by \cite{gao11b}
\begin{equation}
\widetilde{a}_{l}(B) = \widetilde{a}_{\text{bg}l}\left(1-\frac{\Delta_{Bl}}{B-B_{0l}}\right) \;,
\label{eq:gaBparam}
\end{equation}
where $\widetilde{a}_{\text{bg}l}$ is the generalized background scattering length, and $\Delta_{Bl}$ is one measure of the width of the resonance. They can be computed from
\begin{equation}
	\widetilde{a}_l(B = B_{0l} + \Delta_{Bl}) = 0 \;,
\end{equation}
and
\begin{equation}
	\widetilde{a}_{\text{bg}l}= - \widetilde{a}_l(B = B_{0l} + \frac{1}{2}\Delta_{Bl}) \;,
\end{equation}
for an isolated resonance, and can be further refined, when necessary, by fitting $\widetilde{a}_l(B)$ to Eq.~(\ref{eq:gaBparam}) at $B$ fields sufficiently close to $B_{0l}$.

Except for the case of a ``broad'' resonance, the set of parameters $B_{0l}$, $\widetilde{a}_{\text{bg}l}$, and $\Delta_{Bl}$ are generally still insufficient to characterize atomic interaction away from the zero energy \cite{sto05,sim05,koh06,chi10,gao11b}. For a complete characterization of ultracold atomic interaction around a magnetic Feshbach resonance, one generally needs one more parameter, the differential magnetic moment, $\delta \mu_l$,  which is the difference of the magnetic moments in interacting and non-interacting states \cite{chi10,gao11b}. It is given by 
\begin{equation}
	\delta \mu_l = \frac{d\epsilon_l (B)}{d B}\bigg|_{B=B_{0l}},
\end{equation}
and can be found from solutions of $K^{c0}_{l}(\epsilon,B)=0$ at energies $\epsilon$ slight different from zero.

The parameters, $B_{0l}$, $\widetilde{a}_{\text{bg}l}$, $\Delta_{Bl}$, $\delta\mu_l$, along with the $C_6$ coefficient, form one set of parameters that provide a complete description of atomic interaction around a magnetic Feshbach resonance. They constitute the most direct generalization of the $s$ wave description \cite{moe95,koh06,chi10} to other partial waves and are convenient at zero energy (the threshold). The parameters for $^6$Li$^{40}$K in $ba$, $aa$, and $ab$ channels are provided in Table~\ref{tab:LiKAllParam} for all Feshbach resonances that we have found in partial wave $s$ through $h$ and in the field range of 0 G through 1009 G. No $l=2,3,4$ resonances are found in this field range, but a number of $l=5$ resonances are predicted in $ba$, $aa$, and $ab$ channels, in addition to those in $s$ and $p$ waves.

\begin{figure}
\includegraphics[width=\columnwidth]{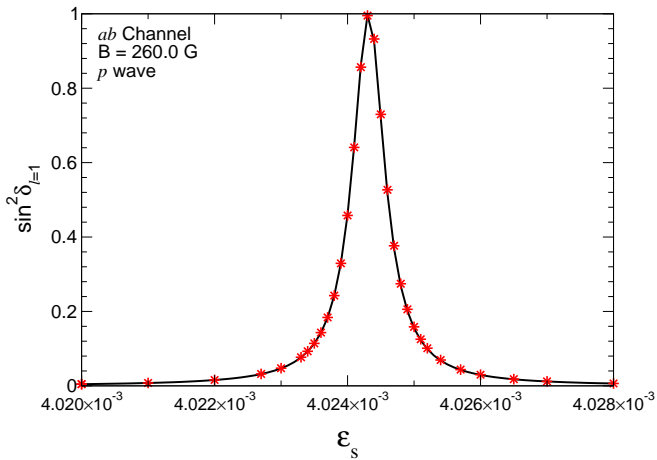}
\caption{(Color online) Plot of $\sin^2\delta_l$ for $p$ wave interaction in $ab$ channel as a function of scaled energy $\epsilon_s=\epsilon/s_E$ at a magnetic field of 260.0 G ($s_E/k_B = 2.490$ mK). The solid line is the results of a numerical calculation from Eq.~(\ref{eq:Kphy}). The symbols are results from the analytic QDT expansion \cite{gao11b} using the parameters specified in Table~\ref{tab:LiKAllParam} for the $p$ wave resonance located at $B_{0l}\approx 259$ G. 
\label{fig:sin2dl}}
\end{figure}

For descriptions of atomic interactions at energies away from the threshold, it is more convenient to use a different set of parameters, $B_{0l}$, $K^{c0}_{\text{bg}l}$, $g_{\text{res}}$, $d_{Bl}$, and $s_E$ (or $C_6$) \cite{gao11b}. They differ from the first set in three parameters that are related by \cite{gao11b}
\begin{equation}
K^{c0}_{\text{bg}l} = \frac{1}{\widetilde{a}_{\text{bg}l}/\bar{a}_{l} -(-1)^l} \;,
\label{eq:Kc0bgga}
\end{equation}
\begin{equation}
g_{\text{res}} = -\frac{\widetilde{a}_{\text{bg}l}/\bar{a}_{l}}{\widetilde{a}_{\text{bg}l}/\bar{a}_{l} -(-1)^l} 
	\left(\frac{\delta\mu_l \Delta_{Bl}}{s_E}\right)\;,
\label{eq:gres}
\end{equation}
\begin{equation}
d_{Bl} = -\frac{\widetilde{a}_{\text{bg}l}/\bar{a}_{l}}{\widetilde{a}_{\text{bg}l}/\bar{a}_{l} -(-1)^l} 
	\Delta_{Bl} \;.
\label{eq:dBl}
\end{equation}
With this set of parameters, which are also provided in Table~\ref{tab:LiKAllParam}, the $K^{c0}_l$ parameter is given by \cite{gao11b}
\begin{equation}
K^{c0}_l(\epsilon_s,B_s) = K^{c0}_{\text{bg}l}
	\left(1+\frac{g_{\text{res}}}{\epsilon_s-g_{\text{res}}(B_s+1)}\right) \;,\\
\label{eq:Kc0Feshs1}	
\end{equation}	
around a Feshbach resonance, where $B_s = (B-B_{0l})/d_{Bl}$. This expression of $K^{c0}_l$, as a function of both energy and magnetic field, goes into the QDT expansion of Ref.~\cite{gao11b} to give an analytic description of atomic interaction in an arbitrary partial wave $l$ and in a $B$ field around $B_{0l}$, not only at the threshold but also in a range of energies away from the threshold. Figure~\ref{fig:sin2dl} depicts an example of the $p$ wave phase shift, more precisely $\sin^2\delta_{l=1}$, around a $p$ wave resonance in $ab$ channel located around $B_{0l}\approx 259$ G. It shows that the results obtained using parameters of Table~\ref{tab:LiKAllParam} and the analytic QDT expansion of Ref.~\cite{gao11b} are in excellent agreement with those computed numerically from Eq.~(\ref{eq:Kphy}). 

Also tabulated in Table~\ref{tab:LiKAllParam} is a derived parameter 
\begin{equation}
\zeta_{\text{res}}\equiv \frac{g_{\text{res}}}{(2l+3)(2l-1)K^{c0}_{\text{bg}l}} \;,
\end{equation}
which distinguishes ``broad'' resonances ($\zeta_{\text{res}}\gg 1$) that follow single-channel universal behaviors from ``narrow'' resonances ($\zeta_{\text{res}}\ll 1$) that deviate strongly from such behaviors \cite{gao11b}. All resonances of $^6$Li-$^{40}$K are found to be narrow, a result that is closely related to the fact that $\Delta a_{l=0}\equiv a^S_{l=0}-a^T_{l=0}$, and thus $\Delta K^c$, is small for $^6$Li-$^{40}$K. It will be interesting to explore other heteronuclear group I systems to look for broad resonances in nonzero partial waves.

\section{Conclusions}

In conclusion, we have presented a MQDT for the interaction of heteronuclear group I atoms in a magnetic field, and have applied it to develop a theory for magnetic Feshbach resonances in such systems. The theory provides a unified and a uniform description of resonances in all partial waves, and enables the description of large number of resonances in terms of very few parameters. The simplified description and the parameters provided by the theory will facilitate further theoretical investigation of two-body, few-body, and many-body systems around magnetic Feshbach resonances, especially those in nonzero partial waves.

For the $^6$Li-$^{40}$K system, we are predicting a number of Feshbach resonances in the $h$ ($l=5$) partial wave. Experimental determination of such high-$l$ resonance positions \cite{kno08}, when compared with the baseline MQDT predictions presented here, will enable much better determinations of both the energy and the partial wave dependences of the short-range parameters such as $\mu^c_S(\epsilon,l)$ and $\mu^c_T(\epsilon,l)$. They will in turn enable more accurate predictions of $^6$Li-$^{40}$K interactions over a much wider range of energies (potentially of the order of 10 K \cite{LYG2014}), both above and below the threshold, and both in and in the absence of the magnetic field.

\begin{acknowledgments}
This work was supported in part by NSF under Grant No. PHY-1306407. BG also thanks Li You, Meng Khoon Tey, and Dajun Wang for helpful discussions and  acknowledges partial support by  NSFC under Grant No.~11328404.
\end{acknowledgments}

\bibliography{bgao,Fesh,atomAtom,qdt,atom,twobody}

\end{document}